\documentclass[twocolumn, twocolappendix]{aastex701}

\usepackage{lipsum} 
\usepackage{caption}
\usepackage{soul}
\renewcommand{\abstractname}{\protect\clearpage ABSTRACT}
\defcitealias{Saha:2026rbf}{LITB III}
\defcitealias{litb2}{LITB II}
\defcitealias{ks1993}{KS93}

\begin{document}

\title{Lensing in the Blue {\MakeUppercase{\romannumeral 4}}: The First Weak Gravitational Lensing Maps from the Stratosphere}

\newcommand\uoftphysics{Department of Physics, University of Toronto}
\newcommand\utias{University of Toronto Institute for Aerospace Studies (UTIAS)}
\newcommand\princeton{Department of Physics, Princeton University}
\newcommand\princetonmae{Department of Mechanical and Aerospace Engineering, Princeton University}
\newcommand\durham{Department of Physics, Durham University}
\newcommand\durhamcfai{Centre for Advanced Instrumentation (CfAI), Durham University}
\newcommand\bps{Boston Public School}
\newcommand\stewardobs{Department of Astronomy, Steward Observatory}
\newcommand\uoftastro{David A. Dunlap Dept. of Astronomy and Astrophysics, University of Toronto}
\newcommand\dunlap{Dunlap Institute for Astronomy and Astrophysics, University of Toronto}
\newcommand\jpl{Jet Propulsion Laboratory, California Institute of Technology}
\newcommand\durhamcea{Centre for Extragalactic Astronomy, Department of Physics, Durham University}
\newcommand\durhamicc{Institute for Computational Cosmology, Department of Physics, Durham University}
\newcommand\imperial{Imperial College London}
\newcommand\oslo{Institute of Theoretical Astrophysics, University of Oslo}
\newcommand\sheffield{Department of Physics and Astronomy, The University of Sheffield}
\newcommand\washu{Department of Physics, Washington University in St.\ Louis}
\newcommand\mcdonnell{McDonnell Center for the Space Sciences, Washington University in St. Louis}
\newcommand\madrid{Departamento de F\'isica Te\'orica,   Universidad Aut\'onoma de Madrid}
\newcommand\nymctu{Institute of Physics, National Yang Ming Chiao Tung University}
\newcommand\northeastern{Department of Physics, Northeastern University}
\newcommand\epfl{Laboratoire d'Astrophysique, EPFL, Observatoire de Sauverny}
\newcommand\casewestern{Department of Physics, Case Western Reserve University}
\newcommand\pltr{Scale AI, Inc.}
\newcommand\starspec{StarSpec Technologies Inc.}
\newcommand\mitaa{Department of Aeronautics and Astronautics, Massachusetts Institute of Technology}
\newcommand\camioa{Institute of Astronomy, University of Cambridge}
\newcommand\camdamtp{Department of Applied Mathematics and Theoretical Physics, University of Cambridge}
\newcommand\camkavli{Kavli Institute for Cosmology, University of Cambridge}
\newcommand\nrcherzberg{National Research Council Canada Herzberg Astronomy and Astrophysics Research Centre}
\newcommand\nyu{Center for Data Science, New York University}
\newcommand\caltech{Department of Physics, California Institute of Technology}
\newcommand\noirlab{Cerro Tololo Inter-American Observatory / NSF NOIRLab}
\newcommand\steward{Department of Astronomy/Steward Observatory, University of Arizona}
\newcommand\wpi{Department of Physics, Worcester Polytechnic Institute}

\newcommand\bls{Boston Latin School}
\newcommand\stanford{Stanford University}
\newcommand\bhs{Belmont High School}
\newcommand\nshs{Newton South High School}

\author[0000-0002-6044-2164]{Sayan Saha}
\affiliation{\northeastern}
\email{sa.saha@northeastern.edu}

\author[0009-0002-2898-7022]{Maya Amit}
\affiliation{\northeastern}
\affiliation{\nyu}
\email{}

\author[0000-0002-9883-7460]{Jacqueline E.\ McCleary}
\affiliation{\northeastern}
\email{j.mccleary@northeastern.edu}

\author[0009-0006-2684-2961]{Georgios N.\ Vassilakis}
\affiliation{\northeastern}
\affiliation{\camioa}
\affiliation{\camdamtp}
\affiliation{\camkavli}
\email{}

\author[0000-0001-5101-7302]{Emaad Paracha}
\affiliation{\uoftphysics}
\email{}  

\author[0000-0002-5273-4634]{Giulia Cerini}
\affiliation{\jpl}
\email{}

\author[0000-0002-3745-2882]{Spencer W.\ Everett}
\affiliation{\caltech}
\email{}

\author[0009-0007-7736-2033]{Leila Ohashi}
\affiliation{\stanford}
\affiliation{\bls}
\affiliation{\northeastern}
\email[]{}  

\author[0000-0002-9378-3424]{Eric M.\ Huff}
\affiliation{\jpl}
\email{}

\author[0009-0004-2523-4425]{Gavin Leroy}
\affiliation{\durhamicc}
\affiliation{\durhamcea}
\email{}

\author[0000-0002-5899-3936]{Leo W.H. Fung}
\affiliation{\durhamicc}
\affiliation{\durhamcea}
\email{}

\author[0000-0002-6085-3780]{Richard Massey}
\affiliation{\durhamicc}
\affiliation{\durhamcea}
\affiliation{\durhamcfai}
\email{}

\author[0000-0002-4214-9298]{Steven J. Benton}
\affiliation{\princeton}
\email{}

\author[0000-0002-3636-1241]{William C. Jones}
\affiliation{\princeton}
\email{}

\author{Thuy Vy T.\ Luu}
\affiliation{\princeton}
\email{}

\author[0000-0002-3937-4662]{Ajay S. Gill}
\affiliation{\nrcherzberg}
\email{}

\author[0000-0002-7600-3190]{Mohamed M.\ Shaaban}
\affiliation{\uoftphysics}
\affiliation{\pltr}
\email{}

\author[0009-0008-5546-890X]{Philippe Voyer}
\affiliation{\princeton}
\affiliation{\princetonmae}
\email{}

\author[0009-0007-7393-2983]{Adam Field}
\affiliation{\wpi}
\email{}

\author[]{Adam Aoua}
\affiliation{\bhs}
\email{}

\author[0000-0003-0259-3148]{Anthony M. Brown}
\affiliation{\durhamcea}
\affiliation{\durhamcfai}
\email{}

\author[0000-0003-0438-2133]{Paul Clark}
\affiliation{\durhamcfai}
\email{}

\author[0000-0002-3162-2107]{Matthew Craigie}
\affiliation{\jpl}
\email{}

\author[0000-0002-2036-2506]{Christopher J.\ Damaren}
\affiliation{\utias}
\email{}

\author[0000-0002-1894-3301]{Tim Eifler}
\affiliation{\steward}
\email{}

\author[0000-0002-6066-6707]{David Harvey}
\affiliation{\epfl}
\email{}

\author[0009-0003-9547-0952]{Eric Habjan}
\affiliation{\northeastern}
\email{}

\author{John W. Hartley}
\affiliation{\starspec}
\email{}

\author{Bradley Holder}
\affiliation{\utias}
\email{}

\author[0000-0003-1974-8732]{Mathilde Jauzac}
\affiliation{\durhamicc}
\email{}

\author[0000-0002-7633-2883]{David Lagattuta}
\affiliation{\durhamicc}
\email{}

\author[0000-0001-7116-3710]{Jason S.-Y.\ Leung}
\affiliation{\imperial}
\email{}

\author[0000-0002-8896-911X]{Lun Li}
\affiliation{\starspec}
\email{}

\author[0000-0002-2036-7008]{Johanna M. Nagy}
\affiliation{\casewestern}
\email{}

\author{C.\ Barth Netterfield}
\affiliation{\uoftphysics}
\affiliation{\uoftastro}
\affiliation{\dunlap}
\email{}

\author[0000-0002-9618-4371]{Susan F.\ Redmond}
\affiliation{\jpl}
\email{}

\author[0000-0002-4485-8549]{Jason D.\ Rhodes}
\affiliation{\jpl}
\email{}

\author[0000-0002-0086-0524]{Andrew Robertson}
\affiliation{\jpl}
\email{}

\author[0000-0002-3884-0983]{L. Javier Romualdez}
\affiliation{\starspec}
\email{}

\author[0000-0001-5612-7535]{J\"urgen Schmoll}
\affiliation{\durhamicc}
\email{}

\author[0000-0002-7542-0355]{Ellen Sirks}
\affiliation{\madrid}
\email{}

\author[0000-0002-6724-833X]{Sut Ieng Tam}
\affiliation{\nymctu}
\email{}

\author[]{Karuna Tarafdar}
\affiliation{\nshs}
\email{}

\author[0000-0002-9740-4591]{Andr\'{e} Z.\ Vitorelli}
\affiliation{\jpl}
\email{}

\author[0000-0001-6455-9135]{Alfredo Zenteno}
\affiliation{\noirlab}
\email{}

\collaboration{all}{SuperBIT collaboration}
\correspondingauthor{Sayan Saha}
\email{sa.saha@northeastern.edu}

\begin{abstract}
The Super-Pressure Balloon-borne Imaging Telescope (SuperBIT) completed its science flight in April 2023, obtaining imaging data for 30 galaxy clusters. In this work, we present the first weak-lensing convergence maps from these observations, using six clusters selected from the 30 targets. We construct background-galaxy catalogs using SuperBIT's three-band photometry and a color-color-based method to separate foreground and background galaxies. Using the calibrated shapes of the selected background galaxies, we reconstruct the weak-lensing convergence field through Kaiser--Squires inversion. We compare the convergence signal-to-noise maps for the six clusters with independent tracers of the mass distribution, namely, red-sequence cluster-member number density maps. These results demonstrate the first successful reconstruction of a weak-lensing signal using imaging data from a sub-orbital astronomical observatory.
\end{abstract}


\section{Introduction} 
Galaxy clusters are among the largest gravitationally bound structures in the Universe and are therefore one of the most powerful probes of cosmic evolution. The abundance of galaxy clusters as a function of their mass and redshift can provide independent constraints on several cosmological parameters, including the dark energy equation of state, the amplitude of matter-density fluctuations $\sigma_8$, and the sum of neutrino masses $\sum m_{\nu}$~\citep{intro1, intro2, intro3, intro4}. Targeted observations of merging cluster systems can also be used to place constraints on alternative dark-matter models \citep{Clowe:2003tk, Robertson:2016xjh, Sirks:2024njj}.

One of the most direct ways to probe the total mass distribution of galaxy clusters is through the gravitational-lensing signatures imprinted on background galaxies. Strong gravitational lenses produce prominent distortions such as arcs and multiple images of background galaxies \citep[][and references therein]{natarajan2024strong}. 
The weak-lensing counterpart produces much smaller, typically sub-percent-level distortions in the two-dimensional light profiles of background galaxies \citep{Bartelmann:1999yn}. Since its first detections~\citep{Wittman:2000tc, Bacon:2000sy, firstwl3, firstwl4}, weak gravitational lensing has become one of the most powerful observational probes to study the dark sector of the Universe.

The reconstruction of weak-lensing signal requires imaging of exceptional quality and meticulous processing to construct accurate catalogs of background galaxy shapes. In the last few decades, many space- and ground-based observatories have successfully measured weak-lensing signals, each with their own advantages and limitations. Ground-based telescopes benefit from the ability to obtain long integration times, allowing them to accumulate sufficient signal from very faint galaxies. However, the unpredictable and spatially varying atmosphere makes it challenging to model the point spread function (PSF) and to accurately resolve and measure the shapes of faint galaxies. Space-based telescopes, on the other hand, provide diffraction-limited imaging but typically have smaller fields of view and require substantially greater resources and costs for deployment and operation.

The Super-Pressure Balloon-borne Imaging Telescope (SuperBIT) was designed to combine some of the advantages of ground- and space-based observations \citep{2018SPIE10702E..0RR, 2020RScI...91c4501R, Gill2024}. During its nearly 45-night science flight in April 2023~\citep{susan2024, Massey2024}, SuperBIT achieved near-diffraction-limited image quality at an altitude of approximately 33~km above sea level, placing it above more than $99\%$ of the Earth's atmosphere. With near-space-like observing conditions, the mission was accomplished at a cost of approximately $10^{-3}$ that of a comparable space-based mission.

This paper is part of the \emph{Lensing in the Blue} (LITB) series within the SuperBIT collaboration. Paper I~\citep{litb1} and Paper II~\citep[hereafter LITB II]{litb2} forecasted the expected weak-lensing efficiency of blue-band observations, while Paper III~\citep[hereafter LITB III]{Saha:2026rbf} presented the processing and calibration of the post-flight SuperBIT imaging data. This paper, Paper IV, presents weak-lensing convergence maps for six of the 30 cluster fields processed in~\citetalias{Saha:2026rbf}, selected based on the strength of their reconstructed weak-lensing signal. We compare the resulting weak-lensing signal with red-sequence cluster-member galaxies, which are independent tracers of the clusters' baryonic matter. To the best of our knowledge, this work presents the first weak-lensing signal reconstruction obtained from the stratosphere, demonstrating the potential of balloon-based observatories as a more economical, and to some degree serviceable alternative to fully space-based missions for future weak-lensing and astrophysical surveys.

The remainder of this paper is organized as follows. In Section~\ref{sec:data}, we provide a brief overview of the data, including the calibrated galaxy shapes and photometry. In Section~\ref{sec:methods}, we describe the methods used in this work, including foreground--background sample separation,  our cluster-member identification scheme, and mass mapping technique. In Section~\ref{sec:results}, we present the weak-lensing convergence maps for our cluster sample, together with comparisons to independent probes such as X-ray emission and the spatial distribution of red-sequence cluster members. Finally, in Section~\ref{sec:conclusion}, we summarize our results and discuss future prospects.

\section{Data}\label{sec:data}
The imaging data obtained during the SuperBIT 2023 science flight have been processed through a dedicated weak-lensing analysis pipeline, resulting in shape catalogs for the detected sources. A detailed description of the data reduction, shape-measurement pipeline, and an extensive suite of systematic tests is presented in \citetalias{Saha:2026rbf}. Here, we provide a brief overview of the final data products relevant to the construction of the weak-lensing convergence maps.

\subsection{Shape Measurement}\label{sec:shapes}

We parameterize the shape of each detected galaxy using the two-component complex ellipticity:
\begin{equation}
\bold{e} = e_1 + i e_2,
\qquad
e_1 = e\cos(2\beta),\quad
e_2 = e\sin(2\beta),
\end{equation}
where $\beta$ is the angle between the x-axis and the major axis. The scalar ellipticity amplitude $e$ is defined as:
\begin{equation}
e = \frac{a-b}{a+b},
\end{equation}
with $a$ and $b$ denoting the semi-major and semi-minor axes, respectively, of the isophotes of the intrinsic, pre-PSF galaxy light profile.

We infer the pre-PSF galaxy parameters using a Bayesian forward-modeling framework implemented with \textsc{ngmix}~\citep{Sheldon:2014vda,ngmix}. 
To correct for observational non-idealities that would otherwise bias the weak-lensing reconstruction, we subsequently calibrate the galaxy shapes using \textsc{Metacalibration}~\citep{Huff:2017qxu,Sheldon:2017szh}.
In this approach, small artificial shears are applied to each galaxy image, and the resulting change in the measured ellipticity is used to estimate its shear response. To first order in the applied shear $\mathbf{\gamma}$,
\begin{equation}
\mathbf{e}(\mathbf{\gamma})
\simeq
\mathbf{e}(0)
+
\mathbf{R}\mathbf{\gamma},
\end{equation}
where $\mathbf{e}(0)$ denotes the ellipticity measured in the absence of an applied shear and  
\begin{equation}
R_{ij}
=
\left.
\frac{\partial e_i}{\partial \gamma_j}
\right|_{\mathbf{\gamma}=0}
\end{equation}
is the \textsc{Metacalibration} shear response matrix. The ensemble-averaged response is measured from the full galaxy sample and used to calibrate the measured ellipticities. The resulting calibrated shapes can then be spatially averaged to estimate the weak-lensing shear field.

For the full SuperBIT 2023 data set, the mean source density after the fiducial selection is $n_{\rm sel}=13.38~\mathrm{arcmin}^{-2}$, with an effective source density~\citep{H12} of $n_{\rm eff}=10.66~\mathrm{arcmin}^{-2}$. These values, however, vary substantially across individual targets as a result of differences in exposure depth, observing conditions, and image quality (see Table 2 of \citetalias{Saha:2026rbf}).  We therefore choose the spatial resolution of the weak-lensing maps, as well as the smoothing kernel size, on a target-by-target basis, taking into account the available source density and the quality of the imaging data.

\subsection{Photometry}\label{sec:photometry}
\begin{figure*}[t]
    \centering
    \includegraphics[width=\linewidth]{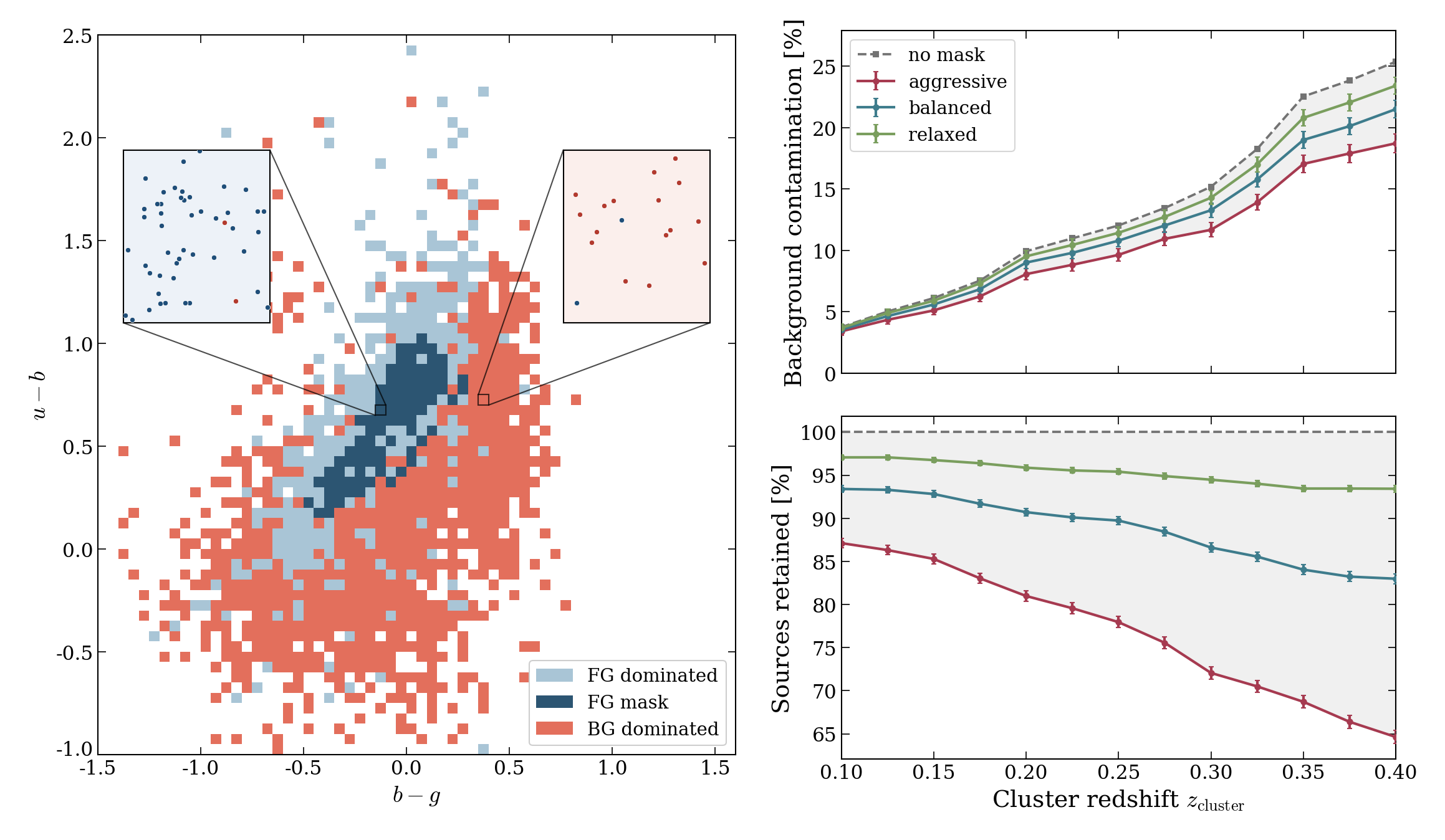}
    \caption{Foreground--background classification in the SuperBIT pipeline. \emph{Left:} An example of the trained pixel mask in the $(u-b)$ versus $(b-g)$ color--color space for a cluster at redshift $z_{\rm cluster}=0.24$. \emph{Right:} Validation of the pixel-mask classifier using independent photometric redshifts from the COSMOS2020 data. The background contamination (top) and retained background-source fraction (bottom) are shown as a function of cluster redshift for three choices of the hyperparameters $(\tau,\eta)$: aggressive (red), balanced (blue), and relaxed (green).}
    \label{fig:pixmask}
\end{figure*}
Taking advantage of the high transmission and low sky background at blue wavelengths~\citep{gill2020}, most of the SuperBIT 2023 imaging data were obtained in three filters at the blue end of the optical spectrum: F400W ($u$), F480W ($b$), and F600W ($g$). The transmission curves of these filters are presented in Figure 2 of \citetalias{litb2}.
We use F480W ($b$) as the detection band and perform forced photometry in F400W ($u$) and F600W ($g$) at the positions of all objects detected in F480W. 
We then calibrate the measured fluxes using the AB-magnitude zero points derived in \citetalias{Saha:2026rbf}.

We correct our galaxy photometry for Galactic dust extinction using the model of \cite{gordon2023}, which provides  wavelength-dependent dust extinction curves $A(\lambda)/A(V)$ parameterized by $R(V)$, the ratio of total-to-selective extinction: 
\begin{equation}
    R(V) = \frac{A(V)}{E(B-V)}.
\end{equation}
We adopt $R(V)=3.1$, a commonly used value representative of diffuse Milky Way dust~\citep{johnson1965}. For each SuperBIT band, we calculate the band-averaged extinction relative to $A(V)$, weighted by the filter transmission:
\begin{equation}
\frac{A_{\rm band}}{A_V} =
\frac{\displaystyle\int S_{\rm band}(\lambda)
\frac{A_\lambda}{A_V}\lambda\mathrm{d}\lambda}
{\displaystyle\int S_{\rm band}(\lambda)\lambda\mathrm{d}\lambda},\label{eqn:extinction_law}
\end{equation}
where $S_{\rm band}(\lambda)$ denotes the transmission of the corresponding SuperBIT filter. We use the \texttt{dustmaps} software~\citep{dustmaps_software} to query the CSFD Galactic dust reddening map \citep{Chiang:2023hyp} at the sky position of each galaxy, calculate
the appropriate $A_{\rm band}$ in each filter from Equation \ref{eqn:extinction_law}, and finally apply the corresponding extinction correction to the galaxies' measured magnitudes. 
\section{Methods}\label{sec:methods}
\subsection{Foreground--Background Separation}\label{sec:fg-bg}
In addition to precise shape measurements, reliable reconstruction of weak-lensing signal requires robust separation of background source galaxies from foreground galaxies. Background source selection is often performed using photometric redshifts, whose accuracy depends on having a sufficient number of independent photometric bands spanning a broad wavelength range to break degeneracies between a galaxy's spectral energy distribution (SED) and its redshift. However, even surveys with four to six broad photometric bands can suffer a 20--30\% catastrophic outlier fraction, driven by color--redshift degeneracies that persist when the available bands do not sufficiently extend into the ultraviolet or infrared \citep{Medezinski:2018}. This limitation is particularly acute for SuperBIT, whose three bands, F400W ($u$), F480W ($b$), and F600W ($g$), occupy a relatively narrow blue-optical wavelength range and overlap substantially in wavelength (Figure 2 of \citetalias{litb2}). Consequently, they provide limited independent information about a galaxy's SED, making photometric redshift estimates highly uncertain. Instead, we adopt a color--color (CC)-based selection scheme that uses the two independent colors available from the three SuperBIT bands: $(u-b)$ and $(b-g)$.

To identify weak-lensing source galaxies, we developed a novel pixel-based classifier in the $(u-b)$ versus $(b-g)$ color--color space. This two-dimensional color space is first divided into a regular grid of ``meta-pixels'', as demonstrated in the left panel of Figure~\ref{fig:pixmask}. 
The pixel-based classifier is then trained using the $(u-b)$ and $(b-g)$ colors of all SuperBIT galaxies with existing external spectroscopic redshift measurements (hereafter, \emph{training objects}). These external spectroscopic redshifts are obtained primarily from the Dark Energy Spectroscopic Instrument (DESI) DR1~\citep{DESI:2025fxa} and the NASA/IPAC Extragalactic Database (NED)\footnote{\url{https://ned.ipac.caltech.edu/}}~\citep{ned}. In total, $4278$ external spectroscopic redshifts are available and used in our pipeline. 
For a cluster at redshift $z_{\rm cluster}$, each training object is labeled as foreground $(z \leq z_{\rm cluster})$ or background $(z > z_{\rm cluster})$. Each meta-pixel is then assigned a foreground \emph{purity}, defined as the fraction of training objects within that pixel that have $(z \leq z_{\rm cluster})$, weighted by the inverse of their mean color variance.
A pixel is flagged as foreground-dominated for a given $z_{\rm cluster}$ when its foreground purity exceeds a threshold $\tau$, provided that it also contains at least $\eta$ training objects; the parameters $\tau$ and $\eta$ serve as the tunable hyperparameters of the classifier. An example of the resulting foreground mask for a cluster at $z_{\rm cluster}=0.24$ is shown in the left panel of Figure~\ref{fig:pixmask}. Of the 352 pixels identified as foreground-dominated by the purity criterion, 88 also satisfy the minimum occupancy requirement and therefore constitute the final mask. 

Once trained, the pixel mask is applied to SuperBIT galaxies without external redshift measurements. Based on their measured colors, objects falling within a foreground-dominated pixel are classified as foreground galaxies, while objects outside the mask are retained as background candidates.

\begin{figure*}[t]
    \centering
    \includegraphics[width=\textwidth]{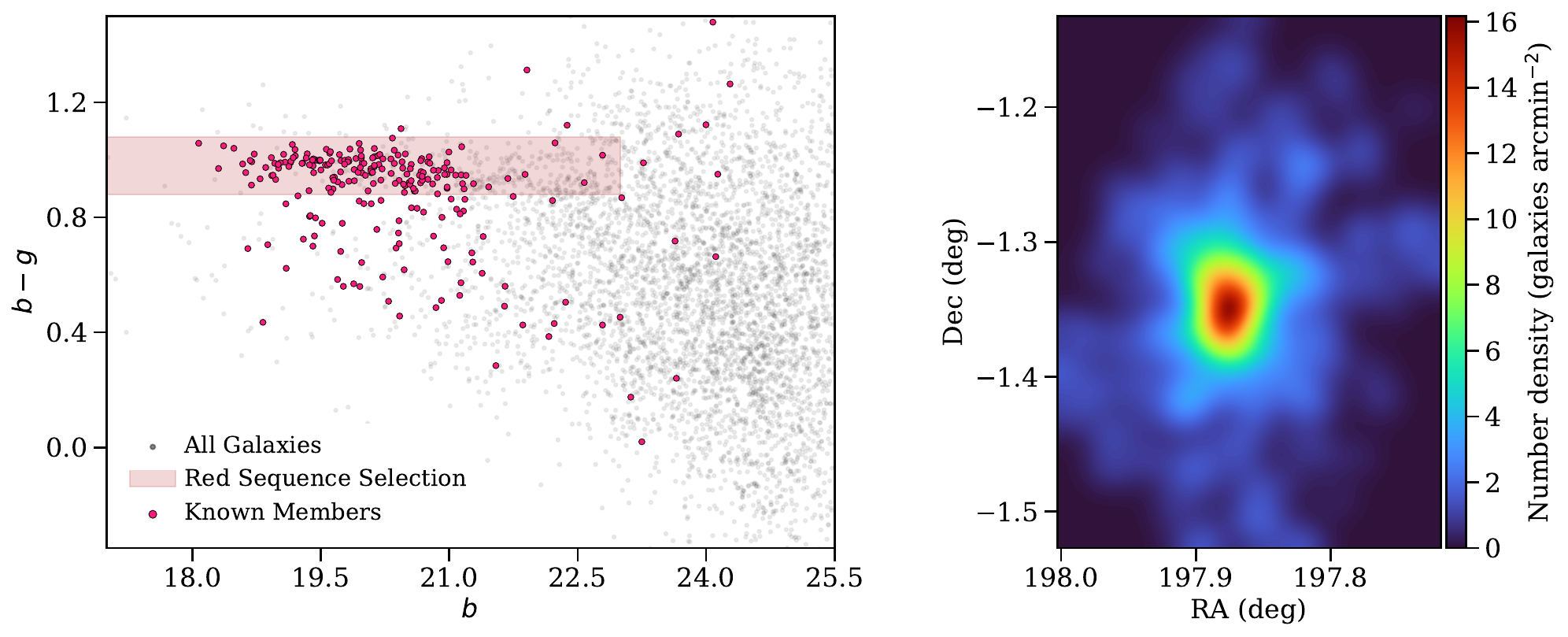}
    \caption{Red-sequence selection for the galaxy cluster Abell~1689. \emph{Left:} Color--magnitude diagram showing $b-g$ color (derived from SuperBIT F480W and F600W magnitudes) versus F480W ($b$)-band magnitude. Grey points denote all galaxies in the field of view; magenta points mark spectroscopically confirmed cluster members. The shaded band indicates the fitted red-sequence selection region; objects falling within this band are classified as red-sequence cluster members. \emph{Right:} Spatial distribution of the selected red-sequence members in RA--Dec coordinates. The colormap shows the projected number density, smoothed with a Gaussian kernel of $0.9\,\mathrm{arcmin}$ FWHM.}
    \label{fig:rs_catcher}
\end{figure*}

We validate the classifier using two SuperBIT pointings in the Cosmic Evolution Survey (COSMOS) field~\citep{cosmos1}, COSMOSk and COSMOS113, for which accurate photometric redshifts are available from the COSMOS2020 catalog~\citep{cosmos2020}. The validation results are shown in the right panel of Figure~\ref{fig:pixmask} for several choices of the hyperparameters $(\tau,\eta)$. We evaluate the classifier's performance using two complementary metrics: the background contamination, defined as the fraction of foreground sources remaining in the retained background sample (upper-right panel), and the background completeness, defined as the fraction of background sources retained after the foreground selection (lower-right panel). We explore a range of hyperparameter choices that trade off contamination against completeness. An aggressive configuration, $(\tau,\eta)=(0.5,1)$, prioritizes low background contamination but removes a larger fraction of genuine background sources. Conversely, a more relaxed configuration, $(\tau,\eta)=(0.9,10)$, prioritizes background completeness at the expense of increased foreground contamination. For the results shown in this work, we adopt an intermediate configuration $(\tau,\eta)=(0.7,10)$, which provides a balanced compromise between background completeness and contamination.

\subsection{Cluster-Member Identification}\label{sec:rs_catching}
Cluster member galaxies provide an important tracer of the baryonic matter distribution within galaxy clusters and share several physical properties that can be exploited for their identification. Galaxies residing in clusters tend to host older stellar populations and generally have lower star-formation rates than galaxies in the field~\citep{dressler1984}. In addition, the buildup of metals through multiple generations of star formation produces a prominent feature in the galaxies' SEDs, known as the 4000~\AA\ break~\citep{Balogh:1999xv}. These features give rise to the well-known red sequence in galaxy color--magnitude diagrams.

The basic procedure of our red-sequence identification algorithm is illustrated in Figure~\ref{fig:rs_catcher}. In the left panel, we show the color--magnitude diagram for galaxies in the field of Abell 1689 ($z_{\rm cluster}=0.1832$). The red points indicate confirmed cluster members, identified as galaxies with external spectroscopic redshifts satisfying $|z-z_{\rm cluster}| < 0.02$. We use these confirmed members to determine the location of the red sequence. We first fit a linear relation to their color--magnitude distribution, iteratively removing outliers through sigma clipping. The fit is flux-weighted, giving higher weight to brighter cluster members with more precise photometric measurements. We then define a red-sequence selection region by allowing a color width of 0.1 mag on either side of the best-fitting relation. We additionally impose a faint-end magnitude cut at $\mathrm{mag}_b=23$. Galaxies falling within this red-sequence region and satisfying the magnitude cut are identified as candidate cluster members. We then examine their spatial distribution in sky coordinates, as illustrated in the right panel of Figure~\ref{fig:rs_catcher}. Based on the color scatter of the known member galaxies, we find that F480W--F600W ($b-g$) provides the smallest scatter among the color combinations considered and therefore adopt this color for red sequence identification. The resulting cluster-member distributions are compared with the clusters' weak-lensing signal in Section~\ref{sec:results}. Our cluster member galaxy identification scheme is described in further detail in Ohashi et al. (in preparation).

\subsection{Mass mapping}\label{sec:ks}
We reconstruct convergence maps from the calibrated galaxy shapes of background galaxies described in Section~\ref{sec:shapes}, following the foreground--background separation described in Section~\ref{sec:fg-bg}, using Kaiser--Squires inversion~\citep[hereafter KS93]{ks1993}. The convergence $\kappa$, which describes the isotropic magnification component of the lensing distortion, can be expressed as the ratio of the projected surface mass density $\Sigma$ to the critical surface density $\Sigma_{\rm crit}$ for the lens--source system:
\begin{equation}\label{eq}
\kappa = \frac{\Sigma}{\Sigma_{\rm crit}},
\qquad
\Sigma_{\rm crit} = \frac{c^2}{4\pi G}\frac{D_{\rm s}}{D_{\rm l}D_{\rm ls}},
\end{equation}
where $D_{\rm s}$, $D_{\rm l}$, and $D_{\rm ls}$ are the observer--source, observer--lens, and lens--source angular-diameter distances, respectively. The Jacobian of the lens mapping can be expressed in terms of the convergence $\kappa$ and the two components of the complex shear, $\gamma = \gamma_1 + i\gamma_2$, both of which are linear combinations of the second-order derivatives of the projected lensing potential $\psi$. Since the intrinsic size distribution of unlensed galaxies is not known, the ensemble-averaged observed galaxy ellipticities provide an estimate of the \textit{reduced} shear, $g = \gamma/(1-\kappa)$. In the weak-lensing regime, where $|\kappa|, |\gamma| \ll 1$, the reduced shear is well approximated by the shear, $g \simeq \gamma$; we therefore take the measured reduced shear as an estimator of $\gamma$ in what follows.

Because both $\kappa$ and $\gamma$ are second derivatives of the same lensing potential, a direct relation between them can be derived (KS93; \citealt{Kaiser:1994jm, Crittenden:2000au, Umetsu2010}) and inverted under the flat-sky approximation \citepalias{ks1993}. This inversion is usually easier to compute in Fourier space:
\begin{equation}\label{eq:ks}
\hat{\kappa}(\mathbf{k}) = \frac{(k_1^2-k_2^2) - 2ik_1k_2}{k_1^2+k_2^2}\,\hat{\gamma}(\mathbf{k}),
\end{equation}
where $\hat{\kappa}(\mathbf{k})$ and $\hat{\gamma}(\mathbf{k})$ are the Fourier transforms of $\kappa(\bold{\theta})$ and $\gamma(\bold{\theta}) = \gamma_1(\bold{\theta}) + i\gamma_2(\bold{\theta})$, respectively, and $k_1$ and $k_2$ are the components of the two-dimensional wave vector $\mathbf{k}$. The real and imaginary parts of the reconstructed convergence correspond to the E- and B-mode components, $\kappa_{\rm E} = \Re(\kappa)$ and $\kappa_{\rm B} = \Im(\kappa)$, respectively \citep[for a review, see][]{Umetsu2020}. The E-mode represents the curl-free gravitational lensing signal, whereas the B-mode provides a useful null test for non-lensing contributions, such as residual observational systematics. We therefore use $\kappa_{\rm B}$ as a null test when choosing selection cuts.

In practice, the post-selection-cut background-galaxy shape catalog is binned into two-dimensional meta-pixels. At each pixel, we estimate the shear from an ensemble average of the galaxy shapes, weighted using the inverse-variance weights of \citetalias{Saha:2026rbf}. Equation~\ref{eq:ks} is then applied to the resulting shear grid, and the reconstructed maps are smoothed with a Gaussian kernel to suppress shape noise. The meta-pixel size and kernel width ($\sigma$) are chosen separately for each target and listed in Table~\ref{tab:cluster_sample}. To quantify the significance of the reconstructed signal, we generate 1000 noise realizations by randomly rotating each galaxy and repeating the mass-mapping procedure. We use the same pixel size and Gaussian smoothing kernel for each realization, and define a signal-to-noise ratio $\mathrm{SNR} = \kappa_{\rm E}/\sigma_\kappa$ at each pixel, where $\sigma_\kappa$ is the pixel-wise standard deviation across the noise realizations. The $\kappa_{\rm B}$ SNR maps are obtained identically. The full procedure is implemented in the open-source \textsc{SMPy} package~\citep{smpy, smpy_zenodo}, which we used for this analysis.

\begin{figure*}[t]
    \centering
    \includegraphics[width=\textwidth]{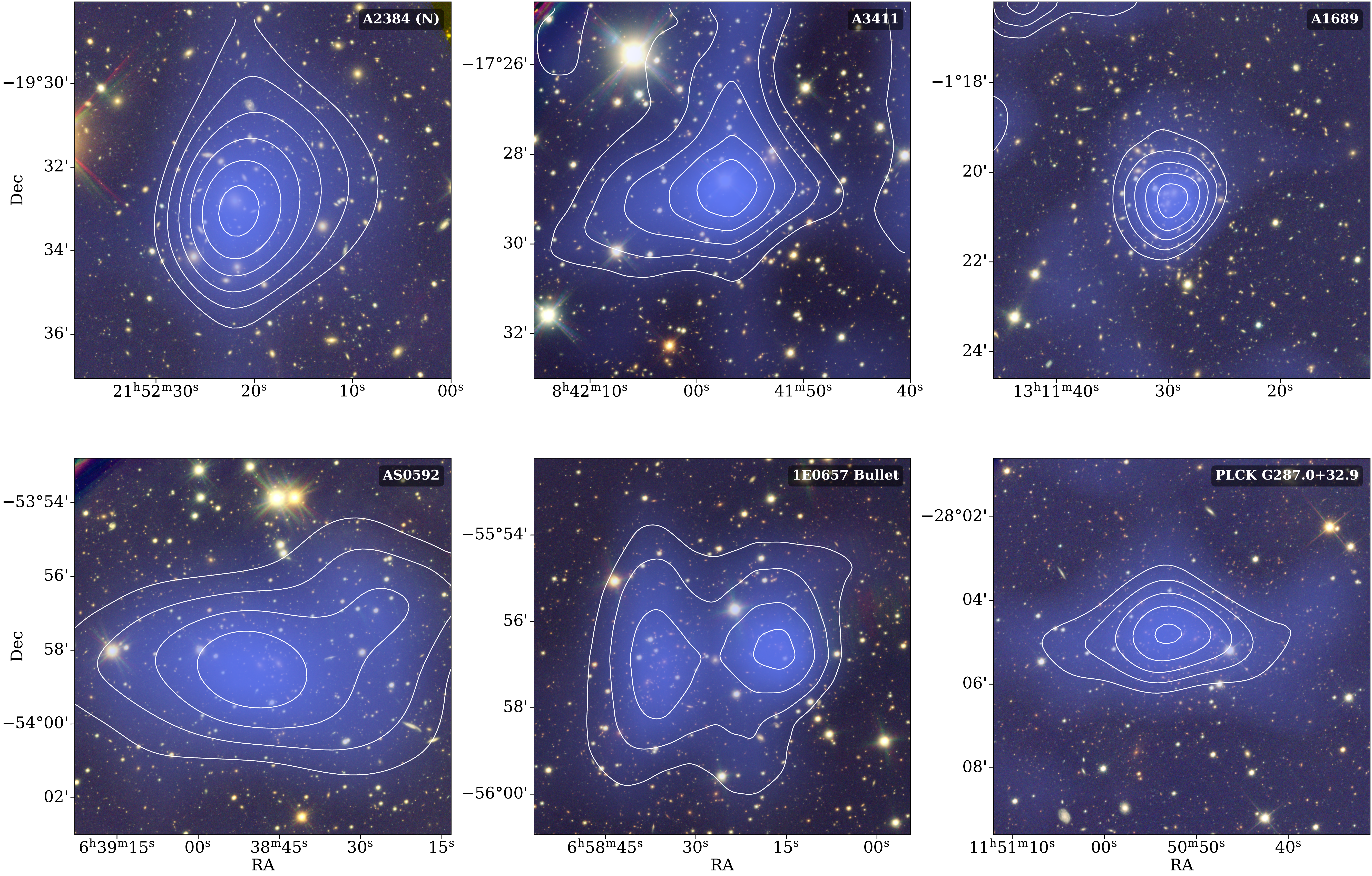}
\caption{Composite optical and mass maps of the six galaxy clusters in our sample. Blue shading indicates the weak-lensing convergence signal, with contours showing the convergence-map SNR starting at $\mathrm{SNR}=2$ in increments of 1. RGB composites were constructed from SuperBIT coadds in the F600W ($g$), F480W ($b$), and F400W ($u$) filters. \textit{Top row, left to right:} A2384 (N), A3411, and A1689. \textit{Bottom row, left to right:} AS0592, 1E\,0657$-$56 (Bullet Cluster), and PLCK~G287.0$+$32.9.}
    \label{fig:cluster_mosaic}  
\end{figure*}
\begin{figure*}[t]
    \centering
    \includegraphics[width=\textwidth]{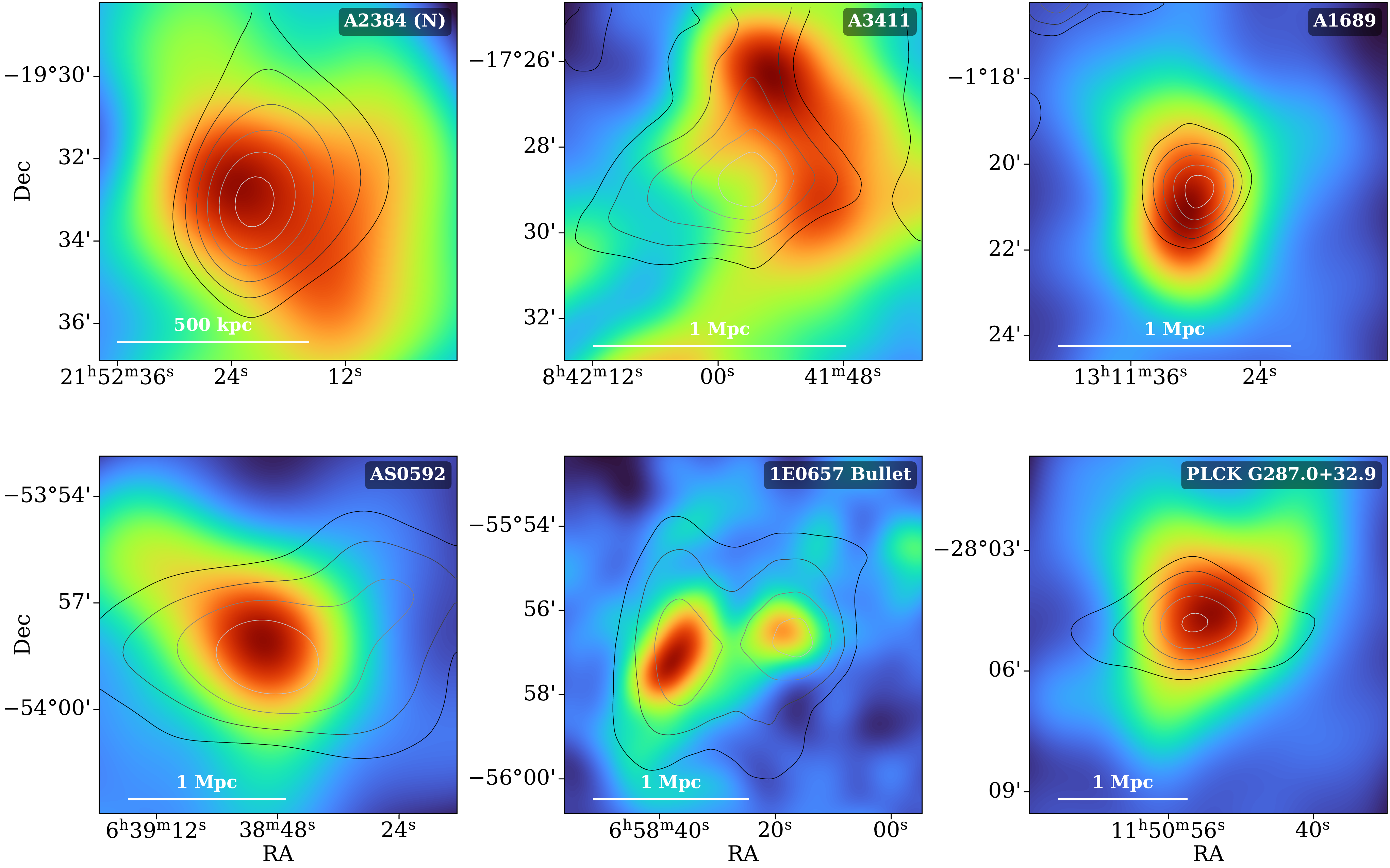}
\caption{Weak-lensing $\kappa_{\rm E}$ SNR contours overlaid on the red-sequence galaxy number-density maps. The upper panels show Abell 3411, Abell 1689, and Abell 2384 (N); lower panels show Abell S0592, PLCK G287.0+32.9, and 1E0657-56 (Bullet). In all panels, the contours start at SNR $=2$ and increase in increments of 1, with progressively lighter contours toward the center representing increasing SNR levels.}

    \label{fig:rs_cluster_mosaic}
\end{figure*}
\section{Results}\label{sec:results}
\begin{deluxetable*}{lcccc c}[t]
\tablecaption{SuperBIT Cluster Sample Information \label{tab:cluster_sample}}
\tablewidth{0pt}
\tablehead{
\colhead{Cluster} & \colhead{Redshift} & \colhead{Selection Cuts} & \colhead{Pixel Size} & \colhead{Kernel Size} & \colhead{SNR$_{\rm peak}$}
}
\startdata
A2384 (N) & 0.094 & $S/N > 10$, $T_{\rm gal}/T_{\rm PSF} > 0.3$ & 0.69\arcmin & 1.70 & 6.16 \\
A3411   & 0.162 & $S/N > 6$, $T_{\rm gal}/T_{\rm PSF} > 0.7$ & 0.47\arcmin & 1.5 & 7.06 \\
A1689   & 0.183 & $S/N > 8$, $T_{\rm gal}/T_{\rm PSF} > 1.0$ & 0.47\arcmin & 1.25 & 6.87 \\
AS0592  & 0.222 & $S/N > 4$, $T_{\rm gal}/T_{\rm PSF} > 0.7$ & 0.65\arcmin & 2.0 & 5.92 \\
1E0657 Bullet       & 0.296 & $S/N > 5$, $T_{\rm gal}/T_{\rm PSF} > 0.5$ & 0.53\arcmin & 1.51 & 5.43 (right lobe), 4.69 (left lobe) \\
PLCK G287.0+32.9  & 0.383 & $S/N > 7$, $T_{\rm gal}/T_{\rm PSF} > 0.5$ & 0.55\arcmin & 1.4 & 5.35 \\
\enddata
\tablecomments{The pixel size refers to the size of each meta-pixel in the reconstructed weak lensing convergence $\kappa_E$ maps. The Gaussian kernel size refers to the $\sigma$ of the smoothing kernel, expressed in units of $\kappa_E$ map meta-pixels.}
\label{tab:top_clusters}
\end{deluxetable*}
In this section, we present the convergence maps reconstructed from the calibrated shapes of background galaxies. We show the results for six clusters for which we obtain the strongest $\kappa_{\rm E}$ detections. As described in \citetalias{Saha:2026rbf}, the baseline source-selection cuts adopted in the pipeline were validated using simulations. However, the quality of individual SuperBIT exposures can vary substantially between targets, and consequently the level of residual systematics in the reconstructed convergence maps can also vary. For each target, we therefore perform a grid search around the baseline selection cuts described in \citetalias{Saha:2026rbf}. We select the combination of cuts that maximizes the $\kappa_{\rm E}$-mode peak SNR while simultaneously minimizing the $\kappa_{\rm B}$-mode SNR in the vicinity of the detected $\kappa_{\rm E}$ peak. The selected targets for this work, their redshifts, peak $\kappa_{\rm E}$ SNRs, and the corresponding source-selection cuts are summarized in Table~\ref{tab:cluster_sample}. We also report the pixel size and Gaussian smoothing kernel used to construct each convergence map. 

Although the SuperBIT 2023 imaging provides a large field of view of approximately $22'\times15.23'$, we do not use the full field for the convergence reconstruction. A roll-stability issue during the observations produced increasingly elliptical PSFs toward the outskirts of the field, resulting in elevated PSF-leakage-related systematics in those regions \citep{Voyer2026}. We therefore crop the images and restrict the analysis to approximately the central $50\%$ of the field of view, where the image quality is more stable.

Figure~\ref{fig:rs_cluster_mosaic} shows the reconstructed convergence maps, with the $\kappa_{\rm E}$ signal represented in blue and its SNR contours overlaid on RGB composite images constructed from the registered SuperBIT coadds in F400W ($u$), F480W ($b$), and F600W ($g$). The corresponding $\kappa_{\rm E}$ and $\kappa_{\rm B}$ SNR maps for each target are presented in Appendix~\ref{app:bmodes}.

We present the comparison of the red-sequence cluster member galaxy distribution with the reconstructed weak-lensing signal  for the six-cluster sample in Figure~\ref{fig:rs_cluster_mosaic}. The background color scale shows the projected density of red-sequence galaxies obtained using the method described in Section~\ref{sec:rs_catching}, with redder colors representing higher galaxy densities and bluer colors representing underdense regions. Overlaid on the red-sequence density maps are contours of the $\kappa_{\rm E}$ signal-to-noise maps. The darker contours toward the outskirts correspond to lower-SNR, while progressively lighter contours trace the higher-SNR regions toward the peaks. The contours start at SNR = 2 and increase in increments of 1. 
Comments for individual clusters are presented below.
\begin{itemize}
    \item \textbf{Abell 2384 (N):} Abell 2384 is a low-redshift merging cluster system at $z=0.094$ comprised of two unequal mass subclusters in a north-south alignment. The image shown here corresponds to the northern and more massive component, A2384 (N); SuperBIT observations also cover the southern subcluster, A2384 (S). The two subclusters are known to be connected via a large bridge of ICM gas, presumably the signature of a recent interaction \citep{2020MNRAS.491.2605P}; little ongoing star formation in disk galaxies near the cluster core provides further evidence of recent interaction \citep{2014A&A...570A..40P}. This cluster is included in the Local Volume Complete Cluster Survey \citep{2024ApJ...974...69F}.  

    \item \textbf{Abell 3411:} 
    Part of a merging cluster system with Abell 3412, Abell 3411 (z=0.169) has been well studied in optical, X-ray, sub-millimeter, and radio wavelengths, and has been the subject of multiple previous weak lensing analyses \citep{2017ApJ...851...46F,2025ApJS..277...28F}. 
    Radio studies of the Abell 3411--Abell 3412 complex revealed a northern extension \citep{2013MNRAS.435..518G} and a giant radio halo \citep{van2013complex}, a feature whose presence generally indicates a recent, massive merger \citep{van2019diffuse}. Abell 3411 is notable because an observed connection between its radio relic and a radio-loud member galaxy
    helped establish that radio galaxies in galaxy clusters provide seed electrons that can be re-accelerated and revived by merger shocks \citep{2017NatAs...1E...5V,zhang2020x}. While Figures~\ref{fig:cluster_mosaic}--\ref{fig:rs_cluster_mosaic} only show the central region of Abell 3411, SuperBIT observations also cover its northern extension and its companion Abell 3412. 
    
    \item \textbf{Abell 1689:} Abell 1689 is another very well-studied galaxy cluster with a radio relic \citep{vacca2011discovery,cuciti2015occurrence} and pre-existing weak lensing studies. It is among the most massive in the observable Universe, with an estimated mass above $10^{15}M_\odot$~\citep{2019ApJ...878...55B}. While azimuthally symmetric on the plane of the sky and dynamically relaxed in its appearance, A1689 is significantly elongated along the line of sight, a fact that has been attributed to a merger event at a very small angle from the line of sight~\citep{andersson2007application}.

    \item \textbf{Abell S0592:} Abell S0592 (also designated SPT-CLJ0638-5358) is a high-mass cluster \citep[$\sim 10^{15} M_\odot$,][]{2019ApJ...878...55B} at $z=0.23$ whose disturbed ICM \citep{mantz2015cosmology, 2022MNRAS.513.3013Y} with at least one known shock \citep{2018MNRAS.476.5591B} and megaparsec-scale diffuse radio emission \citep{2020PASA...37...40W} suggest that it has recently undergone a major merger. Although Abell S0592 has been observed with the Hubble Space Telescope \citep{2021hst..prop16670E}, no gravitational lensing mass maps appear in the literature for it, making ours the first such map for this cluster. 

    \item \textbf{1E 0657-56:} Arguably the most famous galaxy cluster in our sample, 1E 0657-56, better known as the Bullet Cluster, has been used as a dark matter laboratory since the pioneering work of \cite{2004ApJ...606..819M} and \cite{2006ApJ...648L.109C}. Its favorable merger geometry along the plane of the sky means that the Bullet Cluster has been the subject of numerous X-ray \citep[e.g.,][]{2021A&A...650A.104C, 2025A&A...695A.160S}, radio \citep[e.g.,][]{2022A&A...657A..56K,2025ApJ...992...88R}, Sunyaev-Zel'dovich effect \citep[][and references therein]{2024ApJS..272....7B} and gravitational lensing studies, with at least three of the latter released in the last 18 months alone \citep{2025ApJ...987L..15C, 2026A&A...710A.207R, 2026ApJ..1005...28C}, and it continues to serve as a benchmark for hydrodynamical simulations of cluster mergers \citep{Robertson:2016xjh, 2022MNRAS.516.5289M, 2023A&A...673A.131L}. 

    \item \textbf{PLCK G287.0+32.9:} PLCK G287.0+32.9 is the second-most massive cluster in the Planck sample \citep{2016A&A...594A...1P,2020ApJ...892..102L}, with many known strong gravitational lenses surrounding the cluster core \citep{2017ApJ...839L..11Z,d2024powerful}. Prior weak lensing analyses of this cluster appear in \cite{2017ApJ...851...46F} and \cite{2025ApJS..277...28F}. Substructures and radio relics, including ``megahalo'' emission oriented along the merger axis, indicate recent major mergers \citep{bagchi2011discovery, bonafede2014evidence, 2025ApJ...984L..26S,2025ApJ...992...88R}. 
\end{itemize}

\section{Conclusion}\label{sec:conclusion}
We present convergence maps for six galaxy clusters selected from the thirty science targets presented in \citetalias{Saha:2026rbf}. To our knowledge, this work presents the first weak-lensing signal obtained from stratospheric imaging and the first made using blue-band imaging. 
We also make use of the three-band SuperBIT photometry in F400W ($u$), F480W ($b$), and F600W ($g$) to perform foreground--background galaxy separation, an essential step for weak-lensing analyses, as well as to identify cluster-member galaxies using their location in the color--magnitude diagram.

The cluster targets presented here are selected solely on the strength of their reconstructed $\kappa_{\rm E}$ SNR from Kaiser--Squires inversion. A more detailed investigation of the larger SuperBIT cluster sample is underway (G.~Leroy et al., in preparation) and will further develop cluster mass-mapping methods, with particular emphasis on the characterization and mitigation of $\kappa_{\rm B}$ modes, and will explore the physics of the merging clusters.  
Additional future analyses will investigate the dynamical states of the SuperBIT clusters using Fourier-space coherence measurements \citep{Cerini:2022akj, Cerini:2026jpm}.

SuperBIT provided an important proof of concept for balloon-borne, near-space astronomical imaging. Beyond weak-lensing studies, its data can enable a range of science investigations, from galaxy evolution to star formation to the distribution of intracluster dust. SuperBIT's successor mission, GigaBIT \citep{gigabit, gigabit2}, is already in development, with a larger mirror, smaller detector pixels, and an improved guidance system designed to provide deeper imaging with enhanced pointing stability. GigaBIT is forecast to achieve a background-source density of approximately 35 galaxies per arcmin$^2$ for a cluster at redshift $z_{\rm cluster}=0.34$~\citep{ajaythesis}. SuperBIT, and eventually GigaBIT, therefore add a unique capability to NASA's astronomical data legacy by providing near-space-quality imaging at blue wavelengths. 

\begin{acknowledgments}
Authors LO, AA and KT acknowledges the Michael B. Silevitch and Claire J. Duggan Center for STEM Education at Northeastern University and its signature Young Scholars Program. 
Computation for this work was primarily completed using the Explorer Cluster, supported by Northeastern University's Research Computing team. Additional computations were performed by utilizing the CANDIDE cluster at the Institut d'Astrophysique de Paris, which was funded through grants from the PNCG, CNES, DIM-ACAV, and the Cosmic Dawn Center and maintained by Stephane Rouberol. The research was carried out in part at the Jet Propulsion Laboratory, California Institute of Technology, under a contract with the National Aeronautics and Space Administration (80NM0018D0004). The US team acknowledges support from the NASA APRA grant 80NSSC22K0365. Canadian team members acknowledge support from the Canadian Institute for Advanced Research (CIFAR), the Natural Science and Engineering Research Council (NSERC), the Fonds de recherche du Québec (FRQNT), and the Canadian Space Agency (CSA). UK coauthors acknowledge funding from UK STFC (grants ST/V005766/1 and ST/X001075/1).
This research has made use of the NASA/IPAC Extragalactic Database, which is funded by the National Aeronautics and Space Administration and operated by the California Institute of Technology.
\end{acknowledgments}

\software{
NumPy~\citep{numpy},      
SciPy~\citep{scipy}, Astropy~\citep{astropy2013, astropy2018, astropy2022},
Astroquery~\citep{astroquery},
Matplotlib~\citep{Hunter:2007},
\textsc{SMPy}~\citep{smpy, smpy_zenodo}
}

\appendix

\section{$\kappa_{\rm B}$ modes}\label{app:bmodes}
\setcounter{figure}{0}
\renewcommand{\thefigure}{\thesection\arabic{figure}}
\begin{figure*}
    \centering
    \includegraphics[width=\textwidth]{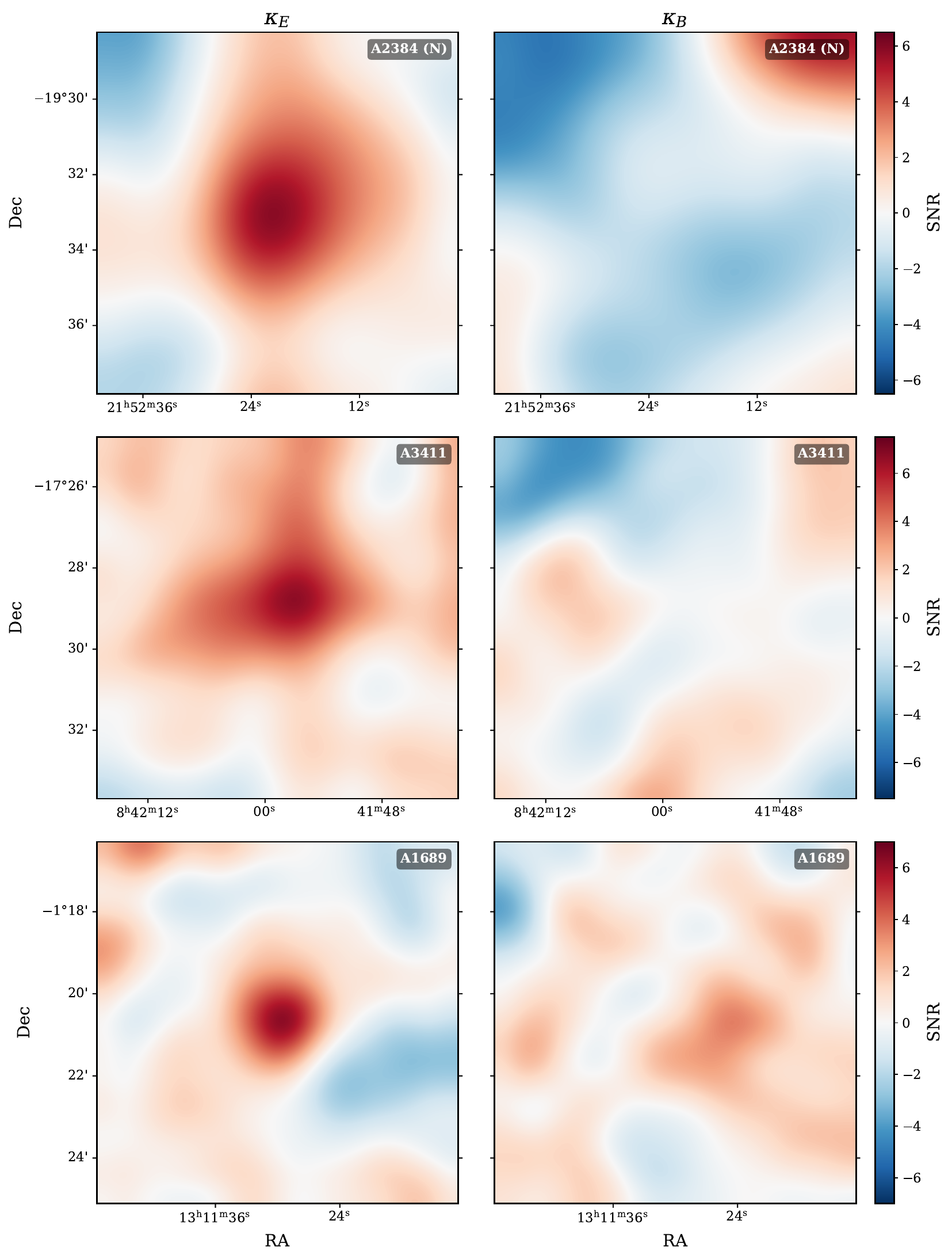}
    \caption{$\kappa_{\rm E}$ and $\kappa_{\rm B}$ SNR maps for the cluster sample.}
    \label{fig:eb_side}
\end{figure*}
\begin{figure*}
    \ContinuedFloat
    \centering
    \includegraphics[width=\textwidth]{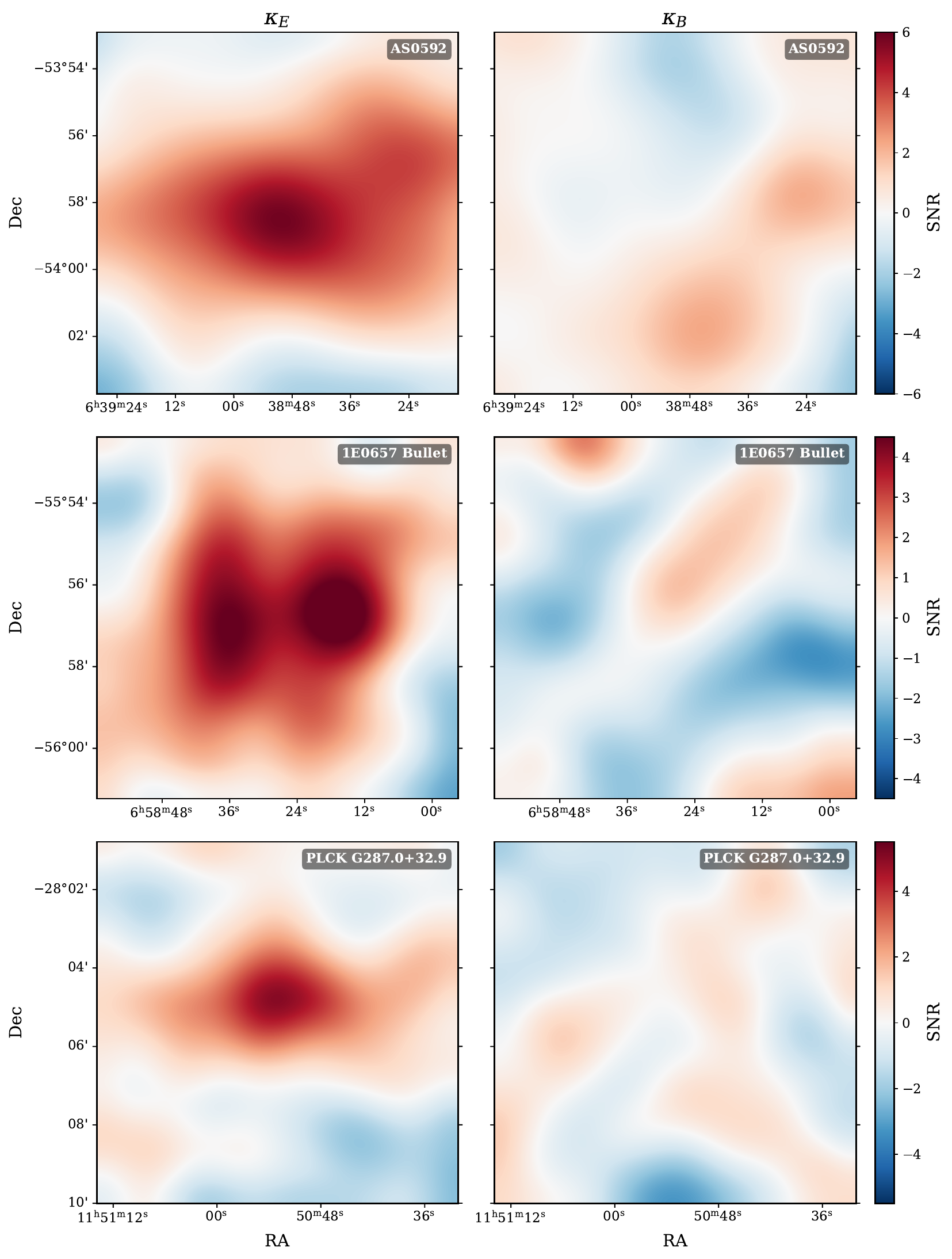}
    \caption{$\kappa_{\rm E}$ and $\kappa_{\rm B}$ SNR maps for the cluster sample (continued).}
    \label{fig:eb_side_cont}
\end{figure*}

In the weak-lensing regime, gravitational lensing by scalar gravitational potentials produces only curl-free distortions of background galaxy shapes, which contribute to the $\kappa_{\rm E}$ mode. By contrast, a non-zero $\kappa_{\rm B}$ mode can arise from several sources, including residual systematics in the shape-measurement pipeline, masking effects, edge effects, and shape noise. In this section, we assess $\kappa_{\rm B}$ modes for the six clusters discussed in Section~\ref{sec:results}. 

In Figures \ref{fig:eb_side} and \ref{fig:eb_side_cont}, we show the $\kappa_{\rm B}$-mode SNR maps obtained using the final selected galaxy catalog for each cluster field. The maps are constructed using the same pixelization and smoothing parameters listed in Table~\ref{tab:cluster_sample}. The definitions of the $\kappa_{\rm E}$ and $\kappa_{\rm B}$ modes, as well as the corresponding SNR estimators, are described in detail in Section~\ref{sec:ks}.
To facilitate visual comparison between the E- and B-mode signals, the color bars of the $\kappa_{\rm E}$- and $\kappa_{\rm B}$-mode SNR maps are set to the same range for each cluster. This allows the relative strength of the B-mode signal with respect to the corresponding E-mode signal to be assessed visually. For all targets presented here, we find that the $\kappa_{\rm B}$ modes are generally $\lesssim 3\sigma$. The exception is A2384 (N), which shows a relatively high $\kappa_{\rm B}$ SNR near the edge of the field of view. A1689 exhibits a $\sim3.5\sigma$ $\kappa_{\rm B}$ signal near the $\kappa_{\rm E}$ peak; however, the $\kappa_{\rm E}$ SNR reaches $6.87$. These maps provide visual evidence that the signals observed in Figures~\ref{fig:cluster_mosaic} and \ref{fig:rs_cluster_mosaic} are genuine and unlikely to be significantly affected by residual systematics. 



\bibliography{main}{}

\bibliographystyle{aasjournalv7}

\end{document}